\documentclass{aa} 
\usepackage[varg]{txfonts}
\usepackage{graphicx}
\usepackage{natbib}
\newcommand{\isotope}[2]{${}^{#1}$#2}
\newcommand{\msun}{\mbox{$\mathrm{M_{\odot}}$}}
\newcommand{\msunb}{\mbox{$\mathrm{M_{\odot}}$} }

\newcommand{\gcc}{\mbox {{\rm g~cm$^{-3}$}}}

\newcommand{\gccb}{\mbox {{\rm g~cm$^{-3}$}} }

\newcommand{\molg}{\mbox {{\rm mol~g$^{-1}$}}}

\begin{document}
   \title{Sensitivity of Type Ia supernovae to electron capture rates}

   \author{E. Bravo
   }

   \institute{E.T.S. Arquitectura del Vall\`es, Universitat Polit\`ecnica de Catalunya, Carrer Pere Serra  
1-15, 08173 Sant Cugat del Vall\`es, Spain\\   
              \email{eduardo.bravo@upc.edu} \label{inst1}}


 
  \abstract{
The thermonuclear explosion of massive white dwarfs is believed to explain at least a fraction of Type Ia supernovae (SNIa). After thermal runaway, electron captures on the ashes left behind by the burning front determine a loss of pressure, which impacts the dynamics of the explosion and the neutron excess of matter. Indeed, overproduction of neutron-rich species such as \isotope{54}{Cr} has been deemed a problem of Chandrasekhar-mass models of SNIa for a long time. I present the results of a sensitivity study of SNIa models to the rates of weak interactions, which have been incorporated directly into the hydrodynamic explosion code. The weak rates have been scaled up/down by a factor ten, either globally for a common bibliographical source, or individually for selected isotopes. In line with previous works, the impact of weak rates uncertainties on sub-Chandrasekhar models of SNIa is almost negligible. The impact on the dynamics of Chandrasekhar-mass models and on the yield of \isotope{56}{Ni} is also scarce. The strongest effect is found on the nucleosynthesis of neutron-rich nuclei, such as \isotope{48}{Ca}, \isotope{54}{Cr}, \isotope{58}{Fe}, and \isotope{64}{Ni}. The species with the highest influence on nucleosynthesis do not coincide with the isotopes that contribute most to the neutronization of matter. Among the last ones, there are protons, \isotope{54,55}{Fe}, \isotope{55}{Co}, and \isotope{56}{Ni}, while the main {\it influencers} are \isotope{54,55}{Mn} and \isotope{55-57}{Fe}, in disagreement with Parikh et al (2013), who found that SNIa nucleosynthesis is most sensitive to the $\beta^+$-decay rates of \isotope{28}{Si}, \isotope{32}{S}, and \isotope{36}{Ar}. An eventual increase in all weak rates on {\it pf}-shell nuclei would affect the dynamical evolution of hot bubbles, running away at the beginning of the explosion, and the yields of SNIa.
  }


   \maketitle
%

\section{Introduction}

Weak interactions on iron-group nuclei (IGN) play a key role in the late stages of stellar evolution. In type II supernovae, first electron captures in the iron core reduce the pressure and start the collapse and, later, beta-decays on neutron-rich nuclei contribute appreciably to the neutrino flux, help to regulate the core temperature \citep{1990auf}, and leave an imprint on the nucleosynthetic yield of the innermost ejected shells \citep{2011lan}. \citet{1994auf} studied the most relevant electron captures in the pre-supernova evolution of massive stars, and identified many iron-group nuclei that may have an influence on the conditions at supernova core collapse. These nuclei were subsequently targets of theoretical studies to refine the associated weak rates. As another example, electron capture supernovae \citep[see][for a recent review]{2018gil} are predicted to be triggered by the transmutation of the late nucleosinthetic products of the most massive intermediate-mass stars with low metallicity progenitors \citep{1980miy}.

The relevance of weak interactions for Type Ia supernovae (SNIa) depends on the progenitor system.
Nowadays, there is debate about the nature of SNIa progenitors, whether they are more or less massive white dwarfs (WD), and whether they are part of a single degenerate or a double degenerate binary system. While all of these scenarios have several points in favour and against \citep[][and references therein]{2012cho,2018jac,2018kil,2019reb}, there are indications that SNIa may be produced by a combination of all of them \citep[e.g.][]{2017sas,2018liu}. In SNIa triggered by the explosion of massive WDs, first, during the pre-supernova carbon simmering phase, electron captures and beta decays drive the equilibrium configuration of the star in response to mass accretion from a companion star and, later, electron captures destabilize the WD and start the dynamical phase of the explosion \citep[e.g.][]{1979yok,2008chm,2017pie}. Early during the explosion, electron captures on IGN reduce the electron pressure and affect the dynamical evolution and the nucleosynthesis of SNIa. On the other hand, SNIa coming from the explosion of sub-Chandrasekhar mass WDs \citep{1994woob,2010fin,2018she} are not expected to be affected by either electron captures o beta-decays during the pre-explosive or explosive phases. However, explosion of WDs more massive than $\sim1$~\msunb may drive the central regions to densities and temperatures high enough that the electron mole number changes significantly. 

The sensitivity of SNIa nucleosynthesis to the weak rates adopted in their modelling has been analysed in a few works, with conflicting results. 
\citet{2000brc} probed the consequences of a global change in the stellar weak rates of IGN owing to new shell model calculations of the Gamow-Teller (GT$^+$) strength distribution of {\it pf}-shell nuclei \citep{1998dea}. These authors found a systematic shift in the centroid of the GT$^+$ strength distribution  and lower stellar weak rates than prior models. They explored the thermonuclear explosions of Chandrasekhar-mass WDs with central densities in the range $(1.7 - 2.1)\times10^9$~\gcc, applying approximate factors to correct for the GT$^+$ centroid offset in those nuclei for which the shell model was not available. The new rates improve the nucleosynthesis, reducing the historical excess production of several neutron-rich isotopes of chromium, titanium, iron, and nickel, among others, in SNIa models. They noticed that protons dominate the neutronization during SNIa explosions and, since their weak rate is not affected by the uncertainties plaguing electron capture rates on IGN, the overall neutronization only depends weakly on these uncertainties. They also identified odd-$A$ and odd-odd nuclei as the largest contributors to the WD neutronization, besides protons. 

On the other hand, \citet{2013pkh} analysed one three-dimensional model of SNIa and the classical one-dimensional W7 model \citep{1986thi}, and found maximal sensitivity of SNIa nucleosynthesis to the electron capture rates of the $\alpha$ elements \isotope{28}{Si}, \isotope{32}{S}, and \isotope{36}{Ar}, whereas electron captures on IGN had little impact on the explosion. Recently, \citet{2016moi} have studied the impact that recent alternatives to the shell model calculations of \citet{1998dea}, motivated by new experimental data on the electron capture rates on \isotope{56}{Ni} and \isotope{55}{Co}, have on SNIa nucleosynthesis. These data call for differences in the GT$^+$ strength distribution, leading to reduced electron capture rates. \citet{2016moi} found that the overall yields of the explosion are affected at most by $2 - 3\%$. Both \citet{2016moi} and \citet{2013pkh} based their sensitivity study on modifying the nuclear rates in nuclear post-processing, meaning that the sensitivity study was decoupled from the supernova hydrodynamics. 

In the present work, I explore the sensitivity of SNIa hydrodynamics and nucleosynthesis to the rates of weak interactions during the dynamic phase of the explosion, using a one-dimensional hydrocode with a large nuclear network that makes unnecessary to post-process the thermodynamic trajectories to obtain the nucleosynthesis. The hydrodynamic model is run for each modification of a weak rate, thus making the calculation of the hydrodynamics and nucleosynthesis consistent. I explore a range of WD central density and mass, and modify the weak rates globally, according to their bibliographic source, and individually for the isotopes that contribute most to the neutronization of the ejected matter.

\section{Explosion models}

\begin{table*}
\caption{Cases studied.} 
\label{t:models} 
\centering 
\begin{tabular}{lllcccccc}
\hline\hline
\noalign{\smallskip}
Model & Class & Type & $M_\mathrm{WD}$ & $\rho_\mathrm{c}$ & $v_\mathrm{def}/v_\mathrm{sound}$ & $\rho_\mathrm{DDT}$ & $K$
& $M(^{56}\mathrm{Ni})$
\\
 & & & $(\msun)$ & $(\gcc)$ & & $(\gcc)$ & ($10^{51}$~erg) & $(\msun)$ \\
\hline
\noalign{\smallskip}
S & sub-$M_\mathrm{Ch}$ & DETO & 1.06 & $4.8\times10^7$ & - & - & 1.32 & 0.664 \\
S+ & sub-$M_\mathrm{Ch}$ & DETO & 1.15 & $9.5\times10^7$ & - & - & 1.46 & 0.894 \\
C2 & $M_\mathrm{Ch}$ & DDT & 1.36 & $2.0\times10^9$ & 0.03 & $2.4\times10^7$ & 1.44 & 0.712 \\
C3 & $M_\mathrm{Ch}$ & DDT & 1.37 & $3.0\times10^9$ & 0.03 & $2.4\times10^7$ & 1.42 & 0.685 \\
C4 & $M_\mathrm{Ch}$ & DDT & 1.38 & $4.0\times10^9$ & 0.03 & $2.4\times10^7$ & 1.41 & 0.666 \\
C5 & $M_\mathrm{Ch}$ & DDT & 1.39 & $5.0\times10^9$ & 0.03 & $2.4\times10^7$ & 1.38 & 0.612 \\
C3\_100 & $M_\mathrm{Ch}$ & DDT & 1.37 & $3.0\times10^9$ & 0.01 & $2.4\times10^7$ & 1.47 & 0.767 \\
C3\_500 & $M_\mathrm{Ch}$ & DDT & 1.37 & $3.0\times10^9$ & 0.05 & $2.4\times10^7$ & 1.45 & 0.697 \\
C3\_1p2 & $M_\mathrm{Ch}$ & DDT & 1.37 & $3.0\times10^9$ & 0.03 & $1.2\times10^7$ & 1.17 & 0.251 \\
C3\_4p0 & $M_\mathrm{Ch}$ & DDT & 1.37 & $3.0\times10^9$ & 0.03 & $4.0\times10^7$ & 1.49 & 0.859 \\
\hline
\end{tabular}
\end{table*}

I have computed SNIa explosion models in spherical symmetry starting from sub-Chandrasekhar (sub-$M_\mathrm{Ch}$) and Chandrasekhar-mass ($M_\mathrm{Ch}$) WDs. The hydrocode integrates simultaneously the hydrodynamics and the nuclear network, and has been described in detail by \citet{2019bra}. Here, I focus on the behaviour of models suitable for normal-luminosity SNIa, characterized by an ejected mass of \isotope{56}{Ni} of the order of $M(^{56}\mathrm{Ni})\sim0.5 - 0.7$~\msun. To this end, I have selected two base models, one describing the central detonation (DETO) of a sub-$M_\mathrm{Ch}$ WD of mass $M_\mathrm{WD}=1.06$~\msun, and the other belonging to the delayed detonation (DDT) of a $M_\mathrm{Ch}$ WD with central density in the range $\rho_\mathrm{c} = (2 - 5)\times10^9$~\gcc, thus extending the range of $\rho_\mathrm{c}$ explored by \citet{2000brc}. The delayed detonation model is characterized by two parameters, the density of transition, $\rho_\mathrm{DDT}$, from a deflagration (subsonic flame propagation near the center of the WD) to a detonation (supersonic combustion wave), and the velocity of the flame during the deflagration phase, $v_\mathrm{def}$, usually prescribed as a fraction of the local sound velocity, $v_\mathrm{sound}$. The configuration parameters are given in Table~\ref{t:models} together with the main explosion properties: kinetic energy, $K$, and $M(^{56}\mathrm{Ni})$. 

Models S, S+, C3, C3\_1p2, and C3\_4p0 in Table~\ref{t:models} are the same as models 1p06\_Z9e-3\_std, 1p15\_Z9e-3\_std, ddt2p4\_Z9e-3\_std, ddt1p2\_Z9e-3\_std, and ddt4p0\_Z9e-3\_std in \citet{2019bra}, and all the details of the code and the models are the same as in that paper unless otherwise stated in what follows. The remaining models reported in this work share the same initial composition as C3 and use the same set of thermonuclear reaction rates. I have explored the impact of different weak rates for various parameters of the DDT and for different initial central densities, but most of the simulations are variations of model C3. Its central density at thermal runaway, $\rho_\mathrm{c}=3\times10^9$~\gcc, is that suggested by some recent studies of the carbon simmering phase \citep{2016mar,2017pie}. A central density as high as $\rho_\mathrm{c}=5\times10^9$~\gcc, or even higher, has been invoked to explain the composition of particular SNIa events \citep{2017dav}, while a value as low as $\rho_\mathrm{c}=2\times10^9$~\gccb provides the best nucleosynthetic match with the Solar System ratios of the IGN \citep{2000brc,2016moi}.

All models in Table~\ref{t:models} share the same set of weak reaction rates, the results of their modification are given later. Next, I explain the details of the weak rates incorporated to the models. 

\subsection{Weak rates}\label{s:weak}

\begin{figure}
\resizebox{\hsize}{!}{\includegraphics{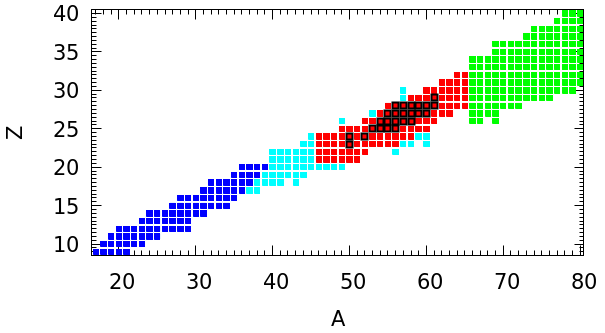}}
\caption{Sources of tabulated weak interaction rates. Colours identifies the source of weak rates on a species: FFN82 in cyan, Oda94 in blue, MPLD00 in red, and PF03 in green. The weak rates on protons, not shown in the graph, are taken from MPLD00. The species with the most influencial individual weak rates (Sect.~\ref{s:indi}) are highlighted with an open black square. All of these belong to the MPLD00 tabulation.
}
\label{f:sources}
\end{figure}

At the high densities characteristic of the central layers of exploding WDs, electron captures are favoured relative to $\beta^+$ decays because the Fermi-Dirac distribution of degenerate electrons allows an enhancement in the electron capture rate, whereas $\beta^+$ decays remain insensitive to the density \citep{2016sar}. The Fermi energy of the electrons depends on density as $E_\mathrm{F}\simeq m_\mathrm{e}\left[\left(\rho_6 Y_\mathrm{e}\right)^{1/3}-1\right]$ with an accuracy better than 98\% for $\rho_6 > 10^3$, where $\rho_6$ is the density in units of $10^6$~\gcc, $Y_\mathrm{e}$ is the electron mole number in \molg, and $m_\mathrm{e}=0.511$~MeV is the mass of the electron in energy units. In SNIa, most of the neutronization takes place on matter that is in a nuclear statistical equilibrium (NSE) state, at temperatures of the order of $T\sim9\times10^9$~K, where the Fermi energy of the electrons is $E_\mathrm{F}\sim 3.5 - 8$~MeV.

Stellar electron capture rates are dominated by allowed GT$^+$ transitions, which allow a change of the nuclear angular momentum from parent to child nuclei by $\Delta J=0,\pm1$. Theory and experiment on ground state GT distribution agree, generally, within a factor $\sim2$ \citep{2012wie}, but when the electron chemical potential (close to the electron Fermi energy) is similar to the reaction $Q$-value, the rates are sensitive to the detailed GT$^+$ distribution, and such distributions cannot be measured from excited states. In theoretical models, the transitions starting from excited states are treated according to the Brink (or Brink-Axel) hypothesis: the GT$^+$ strength distribution from excited states is the same as from the ground state, shifted by the energy of the excited state. However, the applicability of the Brink hypothesis to excited states with low excitation energy is uncertain \citep[e.g.][]{2014mis}.

\citet{1980ful,1982fub,1982ful} calculated weak interaction rates for nuclei in the mass range $A=21-60$ in the IPM approximation (independent-particle model), but the shell model was applied only to {\it sd}-shell nuclei ($A=17 - 40$). Later, \citet{1994oda} revised the rates on {\it sd}-shell nuclei including new relevant experimental information. \citet{1998dea} applied the shell model to the {\it pf}-shell nuclei with experimental data on the GT$^+$ strength distribution: \isotope{51}{V}, \isotope{55}{Mn}, \isotope{54,56}{Fe}, \isotope{59}{Co}, and \isotope{58,60,62,64}{Ni}, most of these even-even nuclei, and predicted the stellar weak rates for other nuclei belonging to the iron group. \citet{2012col} analysed new experimental data on the GT$^+$ strength distribution on {\it pf}-shell nuclei, including those already considered by \citet{1998dea} plus: \isotope{45}{Sc}, \isotope{48}{Ti}, \isotope{50}{V}, and \isotope{64}{Zn}. \citet{2012col} found that the experimental electron captures rates on \isotope{54,56}{Fe} were higher than the theoretical rates \citep{1999cau,2000lmp,2001lan} by as much as a factor two in the conditions of SNIa. \citet{2012fan} analysed the same two isotopes of iron from a purely theoretical point of view, taking into account the uncertainties associated with the different nuclear model parameters, and concluded that the weak rates on these nuclei could change by as much as two orders of magnitude for the whole set of parameters explored. \citet{2013sar,2016sar} revisited the effect of excited states of iron-group nuclei and concluded that thermal excitation of nuclei in SNIa can lead to overall electron capture rates higher as well as lower than those accounting only for transitions from the parent ground state. In either case, the associated uncertainty is similar to that derived from nuclear structure.

In all the models presented in this work, weak interaction 
rates are adopted from, in order of precedence, \citet[][hereafter, MPLD00]{2000gmp}, \citet[][hereafter, Oda94]{1994oda}, \citet[][hereafter, PF03]{2003pru}, and \citet[][hereafter, FFN82]{1982ful}. For instance, if a rate appears both in MPLD00 and in FFN82, the former is the choice. The tables are interpolated following the procedure described in \citet{1985ful}. Figure~\ref{f:sources} shows the sources of each weak rate in the proton number vs baryon number plane. 

\begin{table}
\caption{{\it The neutronizers}: contributors to the overall $\Delta Y_\mathrm{e}$} 
\label{t:neutronizers} 
\centering 
\begin{tabular}{lrrp{35mm}}
\hline\hline
\noalign{\smallskip}
 & $\Delta Y_\mathrm{e,ec}$\tablefootmark{a} & $\Delta Y_{\mathrm{e},\beta^-}$\tablefootmark{b} & Contributing targets 
 \\
 & (\molg) & (\molg) & for e.c. \& $\beta^+$ \tablefootmark{c} \\
\hline
\noalign{\smallskip}
S  & $5.20\times10^{-5}$ & $-3.5\times10^{-9}$ & {\hfill \isotope{60}{Zn}; \isotope{56}{Ni}\hfill} \\
S+ & $1.01\times10^{-4}$ & $-1.2\times10^{-9}$ & {\hfill p; \isotope{60}{Zn}; \isotope{56}{Ni}\hfill}     \\
C2 & $1.11\times10^{-3}$ & $-4.8\times10^{-8}$ & p; \isotope{55}{Co}; \isotope{56}{Ni}; \isotope{54}{Fe}; \isotope{57}{Ni}; \isotope{56}{Co}; \isotope{55}{Fe}; \isotope{58}{Ni} \\
C3 & $2.06\times10^{-3}$ & $-7.8\times10^{-7}$ & p; \isotope{54}{Fe}; \isotope{55}{Co}; \isotope{55}{Fe}; \isotope{56}{Ni}; \isotope{56}{Co}; \isotope{57}{Ni}; \isotope{57}{Co}; \isotope{58}{Ni}; \isotope{59}{Ni}; \isotope{56}{Fe}; \isotope{54}{Mn}; \isotope{50}{Cr} \\
C4 & $2.40\times10^{-3}$ & $-3.3\times10^{-7}$ & p; \isotope{54}{Fe}; \isotope{55}{Fe}; \isotope{55}{Co}; \isotope{56}{Co}; \isotope{56}{Ni}; \isotope{57}{Ni}; \isotope{57}{Co}; \isotope{56}{Fe}; \isotope{58}{Ni}; \isotope{59}{Ni}; \isotope{54}{Mn}; \isotope{50}{Cr}; \isotope{57}{Fe}; \isotope{58}{Co} \\
C5 & $3.13\times10^{-3}$ & $-5.6\times10^{-7}$ & p; \isotope{55}{Fe}; \isotope{54}{Fe}; \isotope{55}{Co}; \isotope{56}{Co}; \isotope{56}{Fe}; \isotope{57}{Co}; \isotope{56}{Ni}; \isotope{57}{Ni}; \isotope{54}{Mn}; \isotope{59}{Ni}; \isotope{58}{Ni}; \isotope{57}{Fe}; \isotope{55}{Mn}; \isotope{50}{Cr}; \isotope{58}{Co}; \isotope{60}{Ni} \\
\hline
\end{tabular}
\tablefoot{
\tablefoottext{a}{Global change in the electron mole number of the WD due to electron captures and positron decays.}
\tablefoottext{b}{Global change in the electron mole number of the WD due to $\beta^{-}$ decays.}
\tablefoottext{c}{Sorted list of species contributing by at least $10^{-5}$~\molg.}
}
\end{table}

Table~\ref{t:neutronizers} gives the overall change in $Y_\mathrm{e}$ as a result of electron captures and $\beta^{+}$ decays, on the one hand, and of $\beta^{-}$ decays, on the other hand, for the sub-$M_\mathrm{Ch}$ models and for the $M_\mathrm{Ch}$ models with different initial central densities. Table~\ref{t:neutronizers} also lists the species that contribute most to the change of $Y_\mathrm{e}$ in each model (from here on, the {\it neutronizers}). As expected, $\beta^{-}$ decay contribution is negligible in all SNIa models. The two sub-$M_\mathrm{Ch}$ models experience a small change of the electron mole number, in spite of their relatively high mass (for a sub-$M_\mathrm{Ch}$ model). The main {\it neutronizers} in the two models are \isotope{60}{Zn} and \isotope{56}{Ni}, and protons as well in the S+ model.

In all $M_\mathrm{Ch}$ models, protons are the main source of neutronization, followed by several isotopes from the IGN, among which there are even-even, odd-odd, and odd-A nuclei. At the lowest $\rho_\mathrm{c}$ explored, the strongest neutronization is provided, besides protons, by \isotope{55}{Co} and \isotope{56}{Ni}, while, for increasing $\rho_\mathrm{c}$, these two species are overtaken by the two iron isotopes \isotope{54,55}{Fe}. Among the IGN reported in Table~\ref{t:neutronizers}, the only rates with direct experimental information about the GT$^{+}$ distribution are \isotope{55}{Mn}, \isotope{54,56}{Fe}, and \isotope{58,60}{Ni}.

\section{Three-dimensional effects}

\begin{figure}
\resizebox{\hsize}{!}{\includegraphics{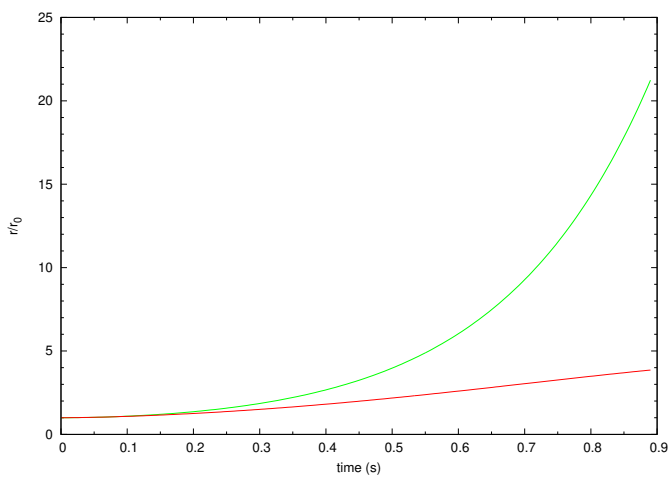}}
\caption{Time evolution of the radial coordinate, $r$, of an incinerated bubble for a central density of the WD of $\rho_\mathrm{c}=3\times10^9$\gcc. The radial coordinate of the bubble is plotted normalized to its initial position, $r_0$, for two cases: standard electron capture rates (green curve) and weak rates scaled up by a factor ten (red curve). 
}
\label{f:bub}
\end{figure}

In three-dimensional models of the thermonuclear explosion of massive WDs, thermal runaway is usually assumed to start in discrete volumes located slightly off-center (bubbles). These bubbles tend to float owing to the expansion caused by the release of nuclear energy, and their density drops off sooner than if they remained at the center and followed the expansion of the whole WD. Consequently the rate of neutronization drops off as well. In models working with standard electron capture rates, the timescale for the bubbles to start rising off the center is $\sim 0.4 - 0.6$~s \citep{2005gse,2018byr}, by which time most of the neutronization has taken place, and the results of the present work are applicable. 

In case the weak rates increased by an order of magnitude, the time taken by the bubbles near the center (therefore, at high density and neutronization rate) would increase sizeably. To estimate this time, I have calculated the dynamic evolution of a hot bubble born near the center of a WD, incorporating the effect of electron captures into the general scheme presented in \citet{2015fis}. In their appendix, \citet{2015fis} wrote a second order differential equation for the evolution of the radial position, $r$, of a hot bubble, accounting for the floatation force and the drag on the bubble:
\begin{equation}\label{eq1}
 \frac{\mathrm{d}}{\mathrm{d}t}\left[\frac{4\pi}{3}R^3\left(\rho_\mathrm{a}+\frac{1}{2}\rho_\mathrm{f}\right)\frac{\mathrm{d}r}{\mathrm{d}t}\right] = 
 \frac{4\pi}{3}R^3\left(\rho_\mathrm{f}-\rho_\mathrm{a}\right)g\,,
\end{equation}
\noindent where $t$ is time since bubble ignition, $R$ is the bubble radius, which is assumed to increase linearly with time (at constant flame velocity), $g=g(r)$ is the local acceleration of gravity, $\rho_\mathrm{f}$ is the local density, and $\rho_\mathrm{a}$ is the density of ashes. To incorporate electron captures in this scheme in a simplified manner, I have assumed that the burning is isobaric, which is valid near the center of the WD, and that the main contribution to pressure is that of a completely degenerate gas of electrons. Then, the pressure, $P$ is a function of the product $\rho Y_\mathrm{e}$ and, to keep it constant, the bubble density changes with time according to 
\begin{equation}
\frac{\dot{\rho}_\mathrm{a}}{\rho_\mathrm{a}}=-\frac{\dot{Y}_\mathrm{e}}{Y_\mathrm{e}}\,.
\end{equation}
Equation~\ref{eq1} can then be integrated numerically, starting from an initial radial coordinate of the bubble, $r_0$. 

Fig.~\ref{f:bub} shows the results for a WD of $\rho_\mathrm{c}=3\times10^9$~\gcc, both for standard electron capture rates and for weak rates increased by a factor ten. With standard electron capture rates, the dynamical evolution of the bubble is similar to the results of complex three-dimensional simulations. However, with increased weak rates the bubble remains near the center, at high density, for nearly a second. This shows that the weak rates have the potential of changing the overall dynamical evolution of $M_\mathrm{Ch}$ models.

\section{Sensitivity of nucleosynthesis to the electron capture rates}

\begin{table*}
\caption{Sensitivity to a bulk change in the rates of a given source.} 
\label{t:sources} 
\centering 
\begin{tabular}{llp{15mm}ccc}
\hline\hline
\noalign{\smallskip}
Model & Source\tablefootmark{a} & {\vskip-0.4cm Scaling \newline factor\tablefootmark{a}} & $\frac{\displaystyle\Delta K}{\displaystyle K}$\tablefootmark{b} & $\frac{\displaystyle\Delta M(^{56}\mathrm{Ni})}{\displaystyle M(^{56}\mathrm{Ni})}$\tablefootmark{b} & Relevant yield ratios\tablefootmark{c} \\
\hline
\noalign{\smallskip}
S & MPLD00 & $\times10$ & 0.002 & -0.021 & \isotope{58}{Ni}$(\times2.2)$; \isotope{62}{Ni}$(\times2.0)$; \isotope{60}{Ni}$(\times0.86)$; \isotope{47}{Ti}$(\times0.41)$ \\
S & MPLD00 & $\times0.1$ & 0.000 & 0.002 & \isotope{47}{Ti}$(\times1.2)$; \isotope{64}{Zn}$(\times1.1)$; \isotope{62}{Ni}$(\times0.89)$; \isotope{58}{Ni}$(\times0.88)$ \\
S+ & MPLD00 - p\tablefootmark{d} & $\times10$ & 0.002 & -0.025 & \isotope{58}{Ni}$(\times1.9)$; \isotope{62}{Ni}$(\times1.6)$; \isotope{60}{Ni}$(\times0.89)$; \isotope{47}{Ti}$(\times0.36)$ \\
C2 & MPLD00 - p & $\times10$ & -0.003 & -0.055 & \isotope{50}{Ti}$(\times27.)$; \isotope{64}{Ni}$(\times23.)$; \isotope{54}{Cr}$(\times14.)$; \isotope{58}{Ni}$(\times1.2)$ \\
C3 & MPLD00 & $\times10$ & -0.030 & -0.130 & \isotope{48}{Ca}$(\times4.9\mathrm{E}5)$; \isotope{64}{Ni}$(\times164.)$; \isotope{54}{Cr}$(\times8.9)$; \isotope{54}{Fe}$(\times0.86)$ \\
C3 & MPLD00 & $\times0.1$ & 0.000 & 0.104 & \isotope{64}{Zn}$(\times1.2)$; \isotope{58}{Ni}$(\times0.69)$; \isotope{48}{Ca}$(\times2.0\mathrm{E-}2)$; \isotope{54}{Cr}$(\times1.7\mathrm{E-}5)$ \\
C3 & MPLD00 - p& $\times10$ & -0.004 & -0.066 & \isotope{48}{Ca}$(\times2.6\mathrm{E}3)$; \isotope{64}{Ni}$(\times34.)$; \isotope{54}{Cr}$(\times3.8)$; \isotope{54}{Fe}$(\times1.01)$ \\
C3 & MPLD00 - p& $\times0.1$ & 0.001 & 0.011 & \isotope{58}{Ni}$(\times1.00)$; \isotope{64}{Zn}$(\times0.99)$; \isotope{54}{Cr}$(\times0.71)$; \isotope{48}{Ca}$(\times0.13)$ \\
C3 & Oda94 & $\times10$ & 0.000 & 0.000 & \isotope{48}{Ca}$(\times1.3)$; \isotope{64}{Ni}$(\times1.1)$; \isotope{54}{Cr}$(\times1.05)$; \isotope{29}{Si}$(\times0.98)$ \\
C3 & Oda94 & $\times0.1$ & -0.001 & -0.003 & \isotope{17}{O}$(\times1.01)$; \isotope{50}{Ti}$(\times0.99)$; \isotope{64}{Ni}$(\times0.99)$; \isotope{48}{Ca}$(\times0.97)$ \\
C3 & FFN82 & $\times10$ & 0.000 & 0.000 & \isotope{48}{Ca}$(\times1.01)$; \isotope{64}{Ni}$(\times1.01)$; \isotope{47}{Ti}$(\times0.998)$; \isotope{33}{S}$(\times0.994)$ \\
C3 & FFN82 & $\times0.1$ & 0.000 & 0.000 & \isotope{22}{Ne}$(\times1.01)$; \isotope{48}{Ca}$(\times0.998)$; \isotope{33}{S}$(\times0.996)$; \isotope{21}{Ne}$(\times0.987)$ \\
C4 & MPLD00 - p& $\times10$ & -0.006 & -0.056 & \isotope{80}{Se}$(\times55.)$; \isotope{48}{Ca}$(\times42.)$; \isotope{64}{Ni}$(\times3.8)$; \isotope{54}{Cr}$(\times2.1)$ \\
C5 & MPLD00 - p& $\times10$ & -0.009 & -0.062 & \isotope{80}{Se}$(\times7.3\mathrm{E}3)$; \isotope{48}{Ca}$(\times14.)$; \isotope{64}{Ni}$(\times2.1)$; \isotope{54}{Cr}$(\times1.6)$ \\
C3\_100 & MPLD00 - p& $\times10$ & -0.004 & -0.037 & \isotope{48}{Ca}$(\times5.1)$; \isotope{64}{Ni}$(\times2.5)$; \isotope{54}{Cr}$(\times1.9)$; \isotope{54}{Fe}$(\times1.05)$ \\
C3\_500 & MPLD00 - p& $\times10$ & -0.004 & -0.074 & \isotope{48}{Ca}$(\times9.3\mathrm{E}3)$; \isotope{64}{Ni}$(\times90.)$; \isotope{54}{Cr}$(\times7.2)$; \isotope{54}{Fe}$(\times4.4)$ \\
C3\_1p2 & MPLD00 - p& $\times10$ & -0.007 & -0.139 & \isotope{48}{Ca}$(\times2.4\mathrm{E}3)$; \isotope{64}{Ni}$(\times33.)$; \isotope{54}{Cr}$(\times3.8)$; \isotope{54}{Fe}$(\times1.01)$ \\
C3\_4p0 & MPLD00 - p& $\times10$ & -0.002 & -0.047 & \isotope{48}{Ca}$(\times2.7\mathrm{E}3)$; \isotope{64}{Ni}$(\times34.)$; \isotope{54}{Cr}$(\times3.8)$; \isotope{54}{Fe}$(\times1.03)$ \\
\hline
\end{tabular}
\tablefoot{
\tablefoottext{a}{The rates from the indicated source were scaled by the factor shown.}
\tablefoottext{b}{Relative change in the final kinetic energy and the mass of \isotope{56}{Ni}, with respect to the values reported for the same model in Table~\ref{t:models}.}
\tablefoottext{c}{Ratio of the final yield of a few selected isotopes with respect to those in the same model with no weak rates scaled. The isotopes reported here are a selection among those with non-negligible ejected mass and whose yield is most affected by the scaling. Ratios larger than 999. are given in exponential format, for example $(\times5\mathrm{E}5)$ means an increment of the yield by a factor $5\times10^5$.}
\tablefoottext{d}{Sources denoted as ``MPLD00 - p'' mean that the rates scaled where those from MPLD00 with the exception of p $\leftrightarrows$ n.}
}
\end{table*}

In this section, I give the results of the sensitivity of the nucleosynthetic yields of the models to the modification of weak rates both globally, applying the same change to all rates coming from a given source (Sect.~\ref{s:sour}), and individually for selected nuclei (Sect.~\ref{s:indi}). In the last case, the sensitivity is measured by the logarithmic derivative, $D_{i,j}$, of the mass ejected of nucleus $i$ with respect to the enhancement factor of the weak rates on species $j$, $f_j$, which is the factor by which these weak rates are scaled at every density and temperature:
\begin{equation}
 D_{i,j} = \frac{\text{d}\log m_{i}}{\text{d}\log f_j} \approx 0.5 \log\left(\frac{m_{i,10}} {m_{i,0.1}}\right)\,,
\label{eqdi}
\end{equation}
where $m_{i,10}$ is the mass of nucleus $i$ ejected for $f_j=10$, and $m_{i,0.1}$ is the corresponding mass when $f_j=0.1$. Just to give a feeling of the meaning of $D_{i,j}$, a value of $D_{i,j}\approx0.3$ means that the abundance of nucleus $i$ doubles for a constant enhancement factor of $f_j=10$. For the same enhancement factor, a value of $D_{i,j}\approx0.05$ implies a relative change in the abundance of a nucleus by $12\%$, and a change of $2\%$ would derive from $D_{i,j}\approx0.01$.

All the sensitivity results are based on the computation of explosion models with individual or global changes in weak rates by factors of either $f_j=10$ or $f_j=0.1$. According to the discussion in Sect.~\ref{s:weak}, it is expected that the weak rates are known with better accuracy, of the order of a factor two. However, as most of the {\it neutronizers} weak rates are not tied by direct experimental results, I explore a slightly larger enhancement factor.

\subsection{Modification of rates by sources}\label{s:sour}

Table~\ref{t:sources} shows the results of the models in Table~\ref{t:models} when all the weak rates tabulated in a given reference are scaled simultaneously by the same factor. In general, the sensitivity to an increase in the weak rates by a factor 10 is much larger than the sensitivity to a decrease in the rates by the same factor. The only exception is the decrease in the rates from MPLD00, including the rates on protons, in model C3, which causes a 10\% increase in the yield of \isotope{56}{Ni} and a decrease in the yield of several neutron-rich nuclei, such as that of \isotope{54}{Cr} by nearly five orders of magnitude. 

The sub-$M_\mathrm{Ch}$ models are very robust against changes in the weak rates. When the MPLD00 rates are increased by a factor ten, the kinetic energy changes just by 0.2\% and the yield of \isotope{56}{Ni} by 2\%. However, it is interesting that the yield of stable nickel doubles that of the model with the standard weak rates.  

Modifying the rates given by Oda94 in $M_\mathrm{Ch}$ models has practically no impact on the supernova explosion dynamics and nucleosynthesis. The modification of the FFN82 rates has no impact, as well, although it should be recalled that I used their rates only when the isotope was not tabulated in the other sources of weak rates. 

The strongest impact of the modification of weak rates is found when all the rates in MPLD00 are increased by a factor ten in model C3, leading to a 3\% decrease in the final kinetic energy and a 13\% decrease in the yield of \isotope{56}{Ni}. In this same run, the yield of \isotope{48}{Ca} increases by nearly six orders of magnitude, that of \isotope{54}{Cr} increases by a factor 8.9, and the yield of \isotope{54}{Fe} decreases by nearly 10\%. However, in this model the proton weak rate was modified by the same factor as the weak rates on IGN, which seems unrealistic since the uncertainty on the proton weak rate is much smaller than those on IGN nuclei. Consequently, I ran the same model with all the rates in MPLD00 modified with the exception of those belonging to protons, for which the standard rate was applied. In this case, the impact is much smaller but still noticeable: the yield of \isotope{56}{Ni} decreases by 6.6\%, that of \isotope{48}{Ca} increases by more than three orders of magnitude, and that of \isotope{54}{Cr} increases by a factor 3.8, while the yield of \isotope{54}{Fe} remains practically unaltered.

When the same modifications are applied to models C2, C4 and C5, in order to explore the effects of different $\rho_\mathrm{c}$, the changes in the kinetic energy and the yield of \isotope{56}{Ni} are similar to those for model C3, but the nucleosynthesis changes in a different way. The higher the initial central density, the less sensitive the yields of \isotope{48}{Ca} and \isotope{54}{Cr} to the modification of the weak rates. The yields of several other isotopes are especially sensitive to the weak rates only when $\rho_\mathrm{c}$ is within a particular range. For instance, this is the case  of the yield of \isotope{80}{Se} at $\rho_\mathrm{c}\sim(4 - 5)\times10^9$~\gcc. 

With respect to the models with different deflagration velocity, models C3\_100 and C3\_500, I find that the nucleosynthesis is increasingly more sensitive to the modification of the weak rates of MPLD00 with increasing $v_\mathrm{def}$. The reason is that, as the deflagration velocity increases, a higher mass is burnt before the white dwarf expands appreciably, then the matter in NSE has more time to capture electrons before weak rates freeze out. The deflagration-to-detonation transition density, $\rho_\mathrm{DDT}$, does not influence the sensitivity of the explosion to weak rates. 
The apparent largest sensitivity of the yield of \isotope{56}{Ni} in model C3\_1p2 simply reflects the small amount of the isotope that is made out of NSE in this model. In absolute terms, the total change in $M(^{56}\mathrm{Ni})$ is very close for all three models with different $\rho_\mathrm{DDT}$ and the same $\rho_\mathrm{c}$, $\Delta M(^{56}\mathrm{Ni})=0.035 - 0.045$~\msun.

\subsection{Modification of individual rates}\label{s:indi}

\begin{table}
\caption{{\it The influencers}: species whose weak rates modification has a large impact on the yield of any other species.} 
\label{t:influencers} 
\centering 
\begin{tabular}{lp{73mm}}
\hline\hline
\noalign{\smallskip}
 & impacted species and $D_{i,j}$ \\
\hline
\noalign{\smallskip}
 \isotope{57}{Fe}: & \isotope{48}{Ca} (0.50); \isotope{64}{Ni} (0.22); \isotope{50}{Ti} (0.09); \isotope{54}{Cr} (0.06); \isotope{67}{Zn} (0.05) \\ \isotope{55}{Mn}: & \isotope{48}{Ca} (0.44); \isotope{64}{Ni} (0.19); \isotope{50}{Ti} (0.07) \\
 \isotope{54}{Mn}: & \isotope{48}{Ca} (0.42); \isotope{4}{He} (0.34); \isotope{64}{Ni} (0.20); \isotope{50}{Ti} (0.09); \isotope{61}{Ni} (0.09); \isotope{54}{Cr} (0.06) \\
 \isotope{56}{Fe}: & \isotope{48}{Ca} (0.42); \isotope{64}{Ni} (0.20); \isotope{50}{Ti} (0.09); \isotope{54}{Cr} (0.07) \\
 \isotope{55}{Fe}: & \isotope{4}{He} (0.40); \isotope{48}{Ca} (0.34); \isotope{64}{Ni} (0.17); \isotope{61}{Ni} (0.12); \isotope{50}{Ti} (0.09); \isotope{54}{Cr} (0.07); \isotope{62}{Ni} (0.06); \isotope{58}{Fe} (0.06) \\
 \isotope{56}{Co}: & \isotope{4}{He} (0.34); \isotope{48}{Ca} (0.11); \isotope{61}{Ni} (0.08); \isotope{64}{Ni} (0.05) \\
 \isotope{58}{Co}: & \isotope{48}{Ca} (0.20); \isotope{64}{Ni} (0.10) \\
 \isotope{56}{Mn}: & \isotope{48}{Ca} (0.18); \isotope{64}{Ni} (0.07) \\
 \isotope{57}{Co}: & \isotope{48}{Ca} (0.17); \isotope{64}{Ni} (0.08) \\
 \isotope{54}{Fe}: & \isotope{48}{Ca} (0.16); \isotope{64}{Ni} (0.08) \\
 \isotope{61}{Ni}: & \isotope{48}{Ca} (0.15); \isotope{64}{Ni} (0.07) \\
 \isotope{59}{Co}: & \isotope{48}{Ca} (0.13); \isotope{64}{Ni} (0.06) \\
 \isotope{60}{Co}: & \isotope{48}{Ca} (0.13); \isotope{64}{Ni} (0.05) \\
 \isotope{50}{V}: & \isotope{48}{Ca} (0.12); \isotope{64}{Ni} (0.06) \\
 \isotope{59}{Ni}: & \isotope{48}{Ca} (0.12); \isotope{64}{Ni} (0.06) \\ 
 \isotope{58}{Fe}: & \isotope{48}{Ca} (0.11) \\
 \isotope{60}{Ni}: & \isotope{48}{Ca} (0.11); \isotope{64}{Ni} (0.05) \\
 \isotope{52}{Cr}: & \isotope{48}{Ca} (0.09) \\
 \isotope{55}{Co}: & \isotope{48}{Ca} (0.07) \\
 \isotope{56}{Ni}: & \isotope{4}{He} (-0.07) \\
 \isotope{50}{Cr}: & \isotope{48}{Ca} (0.06) \\
 \isotope{58}{Ni}: & \isotope{48}{Ca} (0.06) \\
\hline
\end{tabular}
\tablefoot{Listed are those species for which $\left|D_{i,j}\right|\ge0.05$.
}
\end{table}

\begin{table}
\caption{{\it The sensitive}: species most impacted by the modification of individual weak rates.} 
\label{t:sensitive} 
\centering 
\begin{tabular}{lcp{55mm}}
\hline\hline
\noalign{\smallskip}
 & $[$\isotope{A}{$Z$}/\isotope{56}{Fe}$]$\tablefootmark{a} & {\it influencers} and $D_{i,j}$ \\
\hline
\noalign{\smallskip}
 \isotope{4}{He} & -7.2 & \isotope{55}{Fe} (0.40); \isotope{54}{Mn} (0.34); \isotope{56}{Co} (0.34); \isotope{56}{Ni} (-0.07) \\
 \isotope{48}{Ca} & -3.7 & \isotope{57}{Fe} (0.50); \isotope{55}{Mn} (0.44); \isotope{54}{Mn} (0.42); \isotope{56}{Fe} (0.42); \isotope{55}{Fe} (0.34); \isotope{58}{Co} (0.20); \isotope{56}{Mn} (0.18); \isotope{57}{Co} (0.17); \isotope{54}{Fe} (0.16); \isotope{61}{Ni} (0.15) \\
 \isotope{50}{Ti} & 0.47 & \isotope{56}{Fe} (0.09); \isotope{54}{Mn} (0.09); \isotope{55}{Fe} (0.09); \isotope{57}{Fe} (0.09); \isotope{55}{Mn} (0.07); \isotope{58}{Co} (0.04) \\
 \isotope{54}{Cr} & 0.87 & \isotope{55}{Fe} (0.07); \isotope{56}{Fe} (0.07); \isotope{54}{Mn} (0.06); \isotope{57}{Fe} (0.06); \isotope{55}{Mn} (0.04) \\
 \isotope{58}{Fe} & 0.34 & \isotope{55}{Fe} (0.06); \isotope{56}{Fe} (0.05); \isotope{54}{Mn} (0.04) \\
 \isotope{58}{Ni} & -0.16 & \isotope{56}{Ni} (0.04) \\
 \isotope{61}{Ni} & -1.3 & \isotope{55}{Fe} (0.12); \isotope{54}{Mn} (0.09); \isotope{56}{Co} (0.08) \\
 \isotope{62}{Ni} & -0.09 & \isotope{55}{Fe} (0.06) \\
 \isotope{64}{Ni} & -1.1 & \isotope{57}{Fe} (0.22); \isotope{54}{Mn} (0.20); \isotope{56}{Fe} (0.20); \isotope{55}{Mn} (0.19); \isotope{55}{Fe} (0.17); \isotope{58}{Co} (0.10); \isotope{57}{Co} (0.08); \isotope{54}{Fe} (0.08); \isotope{56}{Mn} (0.07); \isotope{61}{Ni} (0.07) \\
 \isotope{64}{Zn} & -2.1 & \isotope{55}{Fe} (0.04) \\
 \isotope{67}{Zn} & -2.5 & \isotope{57}{Fe} (0.05) \\
\hline
\end{tabular}
\tablefoot{Listed are those species for which $\left|D_{i,j}\right|\ge0.045$.
\tablefoottext{a}{$[$\isotope{A}{$Z$}/\isotope{56}{Fe}$]=\log[n($\isotope{A}{$Z$}$)/n_\odot($\isotope{A}{$Z$}$)]-\log[n($\isotope{56}{Fe}$)/n_\odot($\isotope{56}{Fe}$)]$, with $n($\isotope{A}{$Z$}$)$ the number fraction of isotope \isotope{A}{$Z$} in the supernova ejecta, and $n_\odot($\isotope{A}{$Z$}$)$ its abundance in the Solar System \citep{2003lod}.}
}
\end{table}

\begin{figure}
\resizebox{\hsize}{!}{\includegraphics{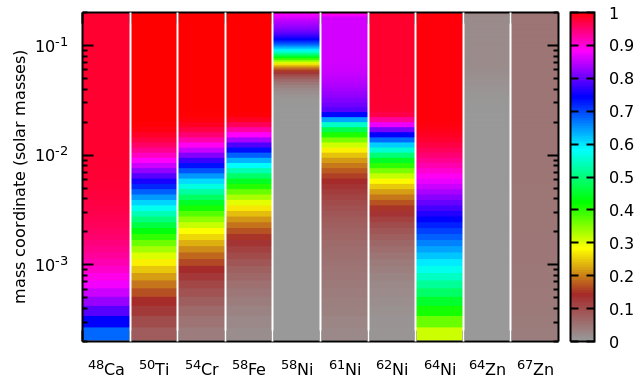}}
\caption{Final distribution of the most sensitive isotopes through the ejecta in model C3. The colour represents the cumulated mass of each isotope, starting from the center of the star, normalized to the total 
ejected mass of the same isotope.
}
\label{f:acum}
\end{figure}

\begin{figure*}
\centering
\includegraphics[width=16.4cm,clip]{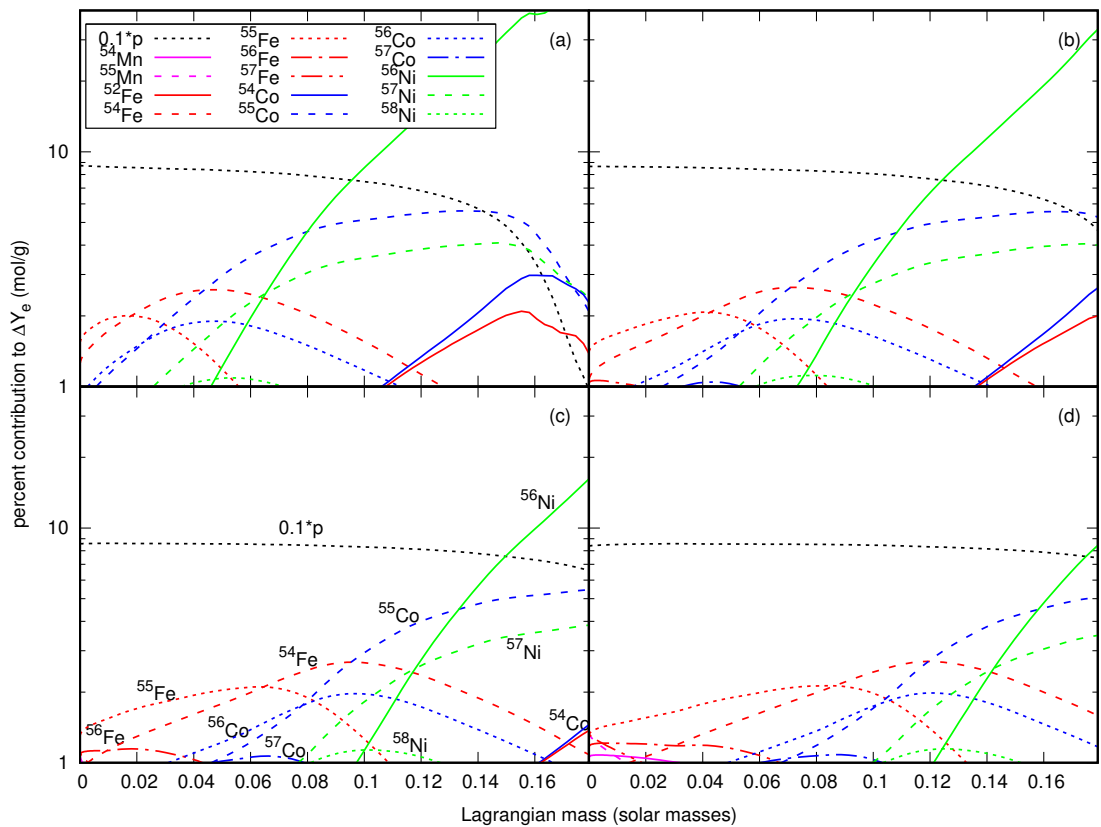}
\caption{Contribution of individual electron captures to the neutronization close to the center as a function of the mass coordinate for a Chandrasekhar-mass model with: (a) $\rho_\mathrm{c}=2\times10^9$~\gccb (model C2), (b) $\rho_\mathrm{c}=3\times10^9$~\gccb (model C3), (c) $\rho_\mathrm{c}=4\times10^9$~\gccb (model C4), and (d) $\rho_\mathrm{c}=5\times10^9$~\gccb (model C5). In panel (c), the curves are labelled with the species name. In all panels, the contribution from protons has been scaled down by a factor ten. 
}
\label{f:capas}
\end{figure*}

Model C3 was rerun with selected weak interaction rates modified by either a factor $f_j=10$ or $f_j=0.1$. The selection criterium for choosing the rates to be modified was that they contributed to the global change of $Y_\mathrm{e}$ in model C3 by at least $5\times10^{-7}$~\molg. The individual rates explored were the electron captures plus $\beta^{+}$ on and the $\beta^{-}$ decays to (all these were changed simultaneously by the same factor) the following list of species: 
\isotope{28}{Si}, \isotope{29,30}{P}, \isotope{32}{S}, \isotope{33,34}{Cl}, \isotope{36}{Ar}, \isotope{50}{V}, \isotope{48,50 - 52}{Cr}, \isotope{50,51,53 - 56}{Mn}, \isotope{52 - 58}{Fe}, \isotope{54 - 60}{Co}, \isotope{55 - 61}{Ni}, \isotope{58 - 61}{Cu}, \isotope{60}{Zn}. Although electron captures on protons dominate the neutronization, I excluded these from the list because of the smaller uncertainty of their tabulated weak rate.  

The kinetic energy in all these models changed by less than 1\% with respect to the reference model C3, and the mass of ejected \isotope{56}{Ni} changed by less than 1\% in all but the models with the rates on \isotope{55,56}{Co} or \isotope{56}{Ni} modified, in which it changed by less than 2\%.

Table~\ref{t:influencers} lists, sorted by $\left|D_{i,j}\right|$, the species (from now on, the {\it influencers}) whose weak rates caused a strong impact on any species with a non-negligible yield. The list is leaded by iron isotopes, then manganese isotopes and cobalt isotopes. To be precise, the top {\it influencer} is \isotope{57}{Fe}, whose $D_{i,j}=0.5$ in relation to the production of \isotope{48}{Ca}, which means that the yield of the last isotope would change by a factor three if the weak rate on \isotope{57}{Fe} increased by a factor ten. \isotope{54,55}{Mn} and \isotope{55,56}{Fe} also stand out among the top {\it influencers}, and all of these impact most the production of \isotope{48}{Ca} and then of \isotope{64}{Ni}. 

These findings contradict the results of \citet{2013pkh}, who identified the rates on \isotope{28}{Si}, \isotope{32}{S}, and \isotope{36}{Ar} as the most influential for the $M_\mathrm{Ch}$ scenario, whereas the weak rates on IGN had little impact on the nucleosynthesis. 
The origin of the discrepancy can be traced back to the treatment of weak rates while matter is in a NSE state. \citet{2013pkh} had overlooked to account for the effect of changes in the weak interactions of the NSE component (private communication). Since the timescales for weak interactions are longer than the WD expansion timescale and beta-equilibrium is not reached in the NSE conditions of SNIa explosions \citep{2000brc}, quantifying the impact of modified individual weak reaction rates while matter is in the NSE state requires a re-computation of the NSE composition at each time step, either of the post-processing calculation or of the hydrodynamic calculation itself, as done in this paper. To confirm that this is the main source of the observed differences, I have rerun model C3 confining the modification of individual weak rates to matter colder than $5\times10^9$~K, to match the NSE criterion applied by \citet{2013pkh}. With this restriction, the most influential rate is the $\beta^+$-decay of \isotope{36}{Ar}, while the impact of the modification of the weak rates on IGN decreases by more than an order of magnitude with respect to the values reported in Table~\ref{t:influencers}.

Within the ten most {\it influencer} electron captures, only the rates on \isotope{55}{Mn} and \isotope{54,56}{Fe} have been determined with the aid of experimental information on the GT$^+$ strength distribution. The existence of low-lying excited states in some of the {\it influencers} contributes to make uncertain their stellar electron capture rates. For instance, the first two excited states of \isotope{57}{Fe} lie at 14.4 and 136.5 keV, to be compared to the thermal energy at the typical temperatures of NSE matter in SNIa, $kT\sim800$~keV. Electron capture from these excited states are favoured by the small difference between their angular momentum and that of the ground state of \isotope{57}{Mn}, which allows a GT$^+$ transition, as opposed to the ground state of \isotope{57}{Fe}. A similar situation occurs for the electron captures on \isotope{56}{Co}.

Table~\ref{t:sensitive} gives the species whose yield is most impacted by the modification of a weak rate on any isotope. The most sensitive are \isotope{48}{Ca} and \isotope{4}{He}, but these two species are of low interest in SNIa from the nucleosynthetic point of view. For instance, in model C3, the ratio of the ejected mass of \isotope{48}{Ca} with respect to the main nucleosynthetic product of SNIa, \isotope{56}{Fe}, normalized to the Solar System ratio is just $2\times10^{-4}$. Of more interest is \isotope{64}{Ni}, which is mainly sensitive to the weak rates on several isotopes of iron and manganese, with $D_{i,j}\sim0.2$. Indeed, if all the weak rates tabulated in MPLD00 minus those involving protons were revised up by a factor 10, \isotope{64}{Ni} would be overproduced with respect to \isotope{56}{Fe} in all $M_\mathrm{Ch}$ models (see Table~\ref{t:sources}). 
The species most overproduced in model C3 (with standard rates) are \isotope{50}{Ti}, \isotope{54}{Cr}, and \isotope{58}{Fe}, partially because of the relatively high initial central density. Their sensitivity to modification of individual weak rates is moderate, $D_{i,j}\sim0.06 - 0.09$.

The pairing of {\it sensitive} and {\it influencer} species can be understood as a result of the spatial coincidence between the regions of synthesis of the former and the regions of maximum impact of the second on the final value of $Y_\mathrm{e}$. Figure~\ref{f:acum} shows, for model C3, the final distribution of the most sensitive species within the WD ejecta. \isotope{48}{Ca} and \isotope{64}{Ni} are synthesized in the innermost few $10^{-3}$~\msunb of the WD, where the contribution of \isotope{57}{Fe} and \isotope{55}{Mn} to the neutronization is maximal, whereas \isotope{50}{Ti}, \isotope{54}{Cr}, \isotope{58}{Fe}, and \isotope{62}{Ni} are made in between mass coordinates $10^{-3} - 10^{-2}$~\msun, where most of the neutronization is provided by p, \isotope{54 - 56}{Fe} and \isotope{54}{Mn}. On the other hand, \isotope{58}{Ni} is made in the mass range $0.06 - 1$~\msun, where most electron captures occur on p, \isotope{56}{Ni} and \isotope{55}{Co}.

Fig.~\ref{f:capas} shows the mass coordinates within which the different species contribute most to the neutronization in the $M_\mathrm{Ch}$ models with different initial central density. At all values of $\rho_\mathrm {c}$, electron captures on protons dominate the neutronization in the central regions of the WD, with decreasing contribution at increasing distance to the center, and electron captures on \isotope{56}{Ni} determine the neutronization at $M\gtrsim0.1 - 0.2$~\msun. The number of species that contribute significantly to the neutronization increases with $\rho_\mathrm{c}$, but their distribution is very similar for all central densities. Therefore, it is to be expected that the sensitivities explored for a $\rho_\mathrm{c}=3\times10^9$~\gccb are representative of models with $\rho_\mathrm{c}$ in the range $(2 - 5)\times10^9$~\gcc.

\section{Summary}

I have assessed the impact of modifications of the weak interaction rates on the nucleosynthesis and other explosion properties of SNIa models. The present work relies on the simulation of one-dimensional models of Chandrasekhar-mass as well as sub-Chandrasekhar mass WD SNIa models through a hydrocode that incorporates a large enough nuclear network that the nucleosynthesis can be obtained directly, and there is no need for nuclear post-processing \citep{2019bra}.

Since many of the arguments in favour of the single-degenerate scenario for SNIa progenitors rely on the detection of neutron-rich nuclei and elements in individual SNIa and their remnants \citep[e.g.][]{2004hoe,2015yam,2018she}, it is of particular interest to test if the explosion of the heaviest sub-Chandrasekhar mass WDs is able to produce the yields required by the observations. And the result is that it is not.
The impact of modifying the electron capture rates in the explosion of sub-Chandrasekhar WDs is small, even for progenitors close to the upper mass limit for WDs made of carbon and oxygen. In particular, model S+, whose progenitor mass is $M_\mathrm{WD}=1.15$~\msun, is unable to produce either the central hole in the distribution of radioactive nickel \citep{2004hoe} or the high nickel to iron mass ratio detected in the supernova remnant 3C397 \citep{2015yam}.

The impact of the explored changes of electron capture rates on the main properties of Chandrasekhar-mass models, namely the final kinetic energy and the mass of \isotope{56}{Ni} synthesized, is also scarce. The yield of \isotope{56}{Ni} may vary by as much as $\sim0.1$~\msun only in case of a global revision of the rates upwards by an order of magnitude, while the maximum variation obtained by changing individual rates is limited to $\sim0.01$~\msun. In comparison, current observational estimates of the amount of \isotope{56}{Ni} needed to power SNIa light curves work with uncertainties of order $\sim0.2$~\msunb \citep{2014sca}, although there are prospects to reduce the error budget to $\sim0.1$~\msunb \citep{2015chi}. 

I have identified three groups of species relevant for assessing the impact of weak rates on SNIa: the {\it neutronizers}, which are those that contribute most to the neutronization of matter, the {\it influencers}, which are those whose eventual weak rate change impacts most the abundance of any species with a non-negligible yield, and the {\it sensitive}, which are the isotopes whose abundance is most impacted by a change in the weak rates. In the Chandrasekhar-mass models explored in this work, the {\it neutronizers} are, besides protons, \isotope{54,55}{Fe}, \isotope{55}{Co}, and \isotope{56}{Ni}. The {\it influencers} do not match the {\it neutronizers} point by point, they are mainly \isotope{54,55}{Mn} and \isotope{55 - 57}{Fe}. Finally, the {\it sensitive} are those neutron-rich nuclei made close to the center of the WD, and which are usually overproduced in Chandrasekhar-mass SNIa models with $\rho_\mathrm{c}>2\times10^9$~\gcc: \isotope{48}{Ca}, \isotope{50}{Ti}, \isotope{54}{Cr}, and \isotope{62,64}{Ni}, in agreement with \citet{2000brc} and \citet{2013pkh}. 

I do not support the claim by \citet{2013pkh} that SNIa nucleosynthesis is most sensitive to the modification of individual $\beta^+$-decay rates of \isotope{28}{Si}, \isotope{32}{S}, and \isotope{36}{Ar}. 
On the other hand, the present results relative to the sensitivity of the nucleosynthesis of SNIa to simultaneous changes in all the weak rates grossly agree with the findings of \citet{2013pkh}. Another effect I have found of increasing globally the weak rates by an order of magnitude is a reduced floatability of hot bubbles near the center, which increases the time available for further electron captures and affects the overall dynamics of the explosion. These results underline the importance of knowing the weak rates with accuracy, at least for Chandrasekhar-mass models. 

\begin{acknowledgements}
I am grateful to Gabriel Mart\'\i nez-Pinedo for providing stellar weak rates tables on a fine grid. Thanks are due to Robert Fisher, for interesting discussions concerning the floatability of hot bubbles, and to Ivo Seitenzahl and Friedrich R\"opke for clarifying the origin of the differences between the present results and those of Parikh et al. The referee, Chris Fryer, has made interesting suggestions to improve the presentation of this paper. This work has been supported by the MINECO-FEDER grant AYA2015-63588-P.
\end{acknowledgements}

\bibliographystyle{aa}
\bibliography{../../ebg}

\end{document}